\documentclass[11pt,showkeys]{revtex4}
\usepackage[utf8x]{inputenc}

\usepackage{amssymb}
\usepackage{latexsym}
\usepackage{amsfonts}
\usepackage{dcolumn}
\usepackage{bm}
\usepackage[usenames]{color}
\usepackage{multirow}
\usepackage{graphicx}
\usepackage{hyperref}
\usepackage{subfig}
\usepackage{booktabs}
\usepackage{xcolor}
\usepackage[normalem]{ulem}
\usepackage{float} 
\usepackage{wrapfig} 
\usepackage{upgreek} 
\usepackage{cancel} 
\usepackage{mathdots} 
\usepackage{mathrsfs} 
\usepackage{phaistos}
\usepackage{wasysym}
\usepackage{hieroglf}
\graphicspath{{figures/}} 

\begin{document}

\title[Complexity measures of high dimensional harmonic and hydrogenic systems]{Complexity measures and uncertainty relations of the high-dimensional harmonic and hydrogenic systems}



\author{N. Sobrino-Coll, D. Puertas-Centeno, I.V. Toranzo and J. S. Dehesa}
\email[]{dehesa@ugr.es}
\affiliation{Departamento de F\'{\i}sica At\'{o}mica, Molecular y Nuclear, Universidad de Granada, Granada 18071, Spain}
\affiliation{Instituto Carlos I de F\'{\i}sica Te\'orica y Computacional, Universidad de Granada, Granada 18071, Spain}

%

\begin{abstract}
In this work we find that not only the Heisenberg-like uncertainty products and the R\'enyi-entropy-based uncertainty sum have the same first-order values for all the quantum states of the $D$-dimensional hydrogenic and oscillator-like systems, respectively, in the pseudoclassical ($D \to \infty$) limit but a similar phenomenon also happens for both the Fisher-information-based uncertainty product and the Shannon-entropy-based uncertainty sum, as well as for the Cr\'amer-Rao and Fisher-Shannon complexities. Moreover, we show that the LMC (López-Ruiz-Mancini-Calvet) and LMC-R\'enyi complexity measures capture the hydrogenic-harmonic difference in the high dimensional limit already at first order.

\end{abstract}

%
%

%

\maketitle

\section{Introduction}

The dimensional scaling has been shown to be a powerful tool to describe numerous features of a great variety of (three dimensional) quantum systems and phenomena pertaining to a wide range of fields from atomic and molecular physics \cite{herschbach1993,tsipis,herschbach96,herschbach_2000,chatterjee} to general quantum mechanics \cite{avery,svidzinsky,dong}, quantum field theory and quantum cosmology \cite{witten,yaffe82,yaffe83,weinberg,kunstaffer} and quantum information \cite{plenio,roven,krenn,bellomo,crann}. It is often possible to approximate the solution of difficult three-dimensional quantum problems  by means of a Taylor series development of similar systems with a non-standard dimensionality (i.e., $D \not= 3$) in powers of $1/D$ and then by using an interpolation or extrapolation procedure. \\

Indeed, guided by the idea that physics at high dimensions is much simpler, Herschbach et al \cite{herschbach1993,tsipis,herschbach_2000} developed a practical method to investigate the electronic structure of atoms and molecules where the dimensional scaling plays the fundamental role. This dimensional scaling method is a very useful strategy to solve the (three-dimensional) finite many-electron problems. This method, which has been mainly used to study the level spectrum and dynamics of Coulomb systems, allows one to solve the associated quantum problem in the \textit{pseudoclassical} ($D\rightarrow\infty$) limit, and then, perturbation theory in $1/D$ is used to have an approximate three-dimensional result, obtaining at times a quantitative accuracy comparable to or better than single-zeta Hartree--Fock calculations~\cite{herschbach1993,tsipis,herschbach96}. Thus, the starting and most interesting point here is the ($D\rightarrow\infty$)-limit, tantamount to $h\to 0$ and/or $m_e\to\infty$ in the kinetic energy, $h$ and $m_e$ being the Planck constant and the electron mass, respectively. This limit is not the same as the conventional classical limit obtained by $h\to 0$ for a fixed dimension~\cite{yaffe82,yaffe83}.\\

It turns out that the pseudoclassical limit of the finite many-electron problem can be exactly computable in a simple way. The electrons of a finite, high $D$-dimensional atom and molecule are confined to harmonic oscillations about the fixed positions attained in the ($D\rightarrow\infty$)-limit. Indeed, in this limit, the electrons of a many-electron system assume fixed positions relative to the nuclei and each other, in~the $D$-scaled space. Moreover, the high-$D$ electronic geometry and energy correspond to the minimum of an exactly known effective potential and can be determined from classical electrostatics for any atom or molecule. Although at first sight, the electrons at rest in fixed locations might seem to violate the uncertainty principle, this is not true because that occurs only in the $D$-scaled space (see, e.g.,~\cite{herschbach_2000}).\\

 The spatial electron delocalization of the main prototypes of the physics of multidimensional quantum systems, the $D$-dimensional hydrogenic system (i.e., a negatively-charged particle moving in a space of $D$ dimensions around a positively charged core which electromagnetically binds it in its orbit) and the $D$-dimensional harmonic system (i.e., a particle moving under the action of a quadratic potential), has been investigated by means of the Heisenberg-like uncertainty measures which are based on the variance and the moments of the quantum probability of the system with order other than two \cite{ray,zozor,toranzo16,aljaber}. Then, uncertainty measures of entropic character \cite{yanez1994,majernik,ghosh,sanchez,assche95,dehesa2001,dehesa1998,dehesa2010,aptekarev2016} have been considered; they are much more appropriate because, contrary to the Heisenberg-like ones, they do not depend on any specific point of the system. 

 Recently these studies have been extended by calculating the dominant term of the Heisenberg-like and R\'enyi-entropy-based uncertainty measures for both the $D$-dimensional hydrogenic and harmonic systems at the quassiclassical border in the two conjugated position and momentum spaces \cite{david1,david2}. It was found that the Heisenberg-like and R\'enyi-entropy-based equality-type uncertainty relations for all of the $D$-dimensional harmonic oscillator states in the pseudoclassical ($D \to \infty$) limit are the same as the corresponding ones for the hydrogenic systems, despite the so different character of the oscillator and Coulomb potentials. In this work, we investigate whether a similar phenomenon also takes place for both the Fisher-information-based uncertainty product and the Shannon-entropy-based uncertainty sum, as well as for various basic (Cr\'amer-Rao, Fisher-Shannon) and generalized (LMC-R\'enyi) complexity measures in the two position and momentum spaces. \\
   
The structure of the paper is the following. First, in Section \ref{sec1} we define and briefly discuss the spreading properties of Heisenberg, entropy and complexity types of a $D$-dimensional probability density which we need for the rest of this work. In Section \ref{sec2} we give the probability densities of the $D$-dimensional hydrogenic and harmonic systems in both position and momentum spaces. In Section \ref{sec3} we calculate and compare not only the position and momentum Heisenberg-like and R\'enyi-entropy-based uncertainty relations but also the  Fisher-information-based uncertainty product and the Shannon-entropy-based uncertainty sum of the $D$-dimensional hydrogenic and harmonic systems. Then, in Section \ref{sec4} we determine and compare
 the basic and extended complexity measures defined above in both position and momentum spaces. Finally, some conclusions and open problems are given.

\section{Entropy and complexity measures of a $D$-dimensional quantum density} 
\label{sec1}

The physical and chemical properties of these systems are controlled \cite{parr} by means of the spatial delocalization or spreading of the single-particle density $\rho (\vec{r}), \vec r \in \Delta \subseteq \mathbb{R}^D$ defined as 
\begin{equation}\label{eq:density}
\rho(\vec{r}):=\sum_{\sigma_1,\sigma_2,...,\sigma_n} \int_{\Delta} \left| 
\Psi (\vec{r},\vec{r}_2,...,\vec{r}_n;\sigma_1,\sigma_2,...,\sigma_n)
  \right|^2 d\vec{r}_2...d\vec{r}_n
\end{equation}
where $\Psi(\vec{r}_1,\vec{r}_2,...,\vec{r}_n;\sigma_1,\sigma_2,...,\sigma_n)$ represents the wave function of the $D$-dimensional $n$-particle system, and $\vec{r}=(x_1,x_2,...,x_D), \sigma_i \in ({-\frac{1}{2}},{\frac{1}{2}}),$ and $(r_i,\sigma_i)$ denote the position-spin coordinates of the $i$th-particle, which is assumed to be normalized and antisymmetrized in the pairs $(\vec{r}_i,\sigma_i)$. The power moments or radial expectation values $\left\{\left< r^\alpha\right> \right\}$ and the entropic moments $\left\{ W_\alpha[\rho] \right\}$ of the density  defined by
\begin{equation}
\label{eq:4}
\langle r^{\alpha}\rangle :=\int_{\Delta} r^{\alpha}\rho(\vec{r})\, d\vec{r}\quad and \quad W_q\left[\rho\right] := \left< \rho^{q-1} \right>= \int_{\Delta} \left[ \rho(\vec{r})\right]^q d \vec{r},
\end{equation}
respectively, provide two different (but equivalent) ways to characterize the density $\rho (\vec{r})$ according to the power-moment and entropic-moment problems of Hamburger type \cite{akhiezer,romera2001}. Moreover they describe numerous physical quantities of the system \cite{gadre91,thakkar04,dehesa2012}.\\ 

The internal disorder (spatial spreading) of the multidimensional quantum systems can be quantified by entropy and complexity measures (which are closely related to the entropic moments of $\rho (\vec{r})$) in a much better way than by the Heisenberg-like measures which are linked to the power moments (radial expectation values) of the single-particle probability density $\rho (\vec{r})$, mainly because the latter ones depend on some specific spatial point of the system; so, e.g. the variance $V[\rho] = \langle \vec{r}^{2}\rangle - \langle \vec{r} \rangle^{2}$ depends on the centroid. The entropy measures (Shannon entropy, disequilibrium, R\'enyi entropy, Fisher information) not only quantify a spreading facet of the electronic density, but also characterize a great deal of fundamental and/or experimentally accessible energetic quantities of the system as shown by the density functional theory. The Shannon entropy \cite{shannon,gyfto} and the disequilibrium (also called informational energy \cite{onicescu}), defined by
\begin{equation}
\label{SHA}
S[\rho] := -\int_{\Delta} \rho(\vec{r})\ln \rho(\vec{r})  \, d\vec{r},\quad  and \quad \mathcal{D}[\rho] := \int_{\Delta} \rho(\vec{r})^{2}\, d\vec{r},
\end{equation}
measure the total spatial spreading of the electronic charge and the separation with respect to the equiprobability, respectively. The corresponding  quantities for the momentum-space probability density $\gamma(\vec{p})$ will be denote by $S \left[ \gamma \right]$ and $\mathcal{D}[\gamma]$, respectively. Other global spreading facets of the position density $\rho(\vec r)$ are given by the monoparametric R\'enyi entropies \cite{renyi_61,leonenko}, $R_q[\rho],$ 
\begin{equation}
\label{REN}
R_{q}[\rho]:=\frac{1}{1-q}\, \ln \int_{\Delta}\rho(\vec{r})^{q}d\vec r, \quad q > 0, \quad q \not= 1.
\end{equation}
Note that these quantities include the Shannon entropy, the disequilibrium and the Tsallis entropies $T_{q}[\rho] =  \frac{1}{q-1} (1-\int_{\mathbb{R}^3} [\rho(\vec{r})]^{q})$, since $S[\rho] = \lim_{q\rightarrow 1} R_{q}[\rho]$, $\mathcal{D}[\rho] = \exp(-R_{2}[\rho])$ and $T_{q}[\rho] = \frac{1}{1-q}[e^{(1-q)R_{q}[\rho]}-1].$ The R\'enyi entropies provide the most relevant canonical class of uncertainty measures \cite{portesi,jizba2016} and characterize different quantities of the system depending on the $q$-parameter  such as the Dirac exchange energy ($q=4/3$), the Thomas-Fermi energy ($q=5/3$), the average density ($q=2$), ... (see e.g., \cite{dehesa2010,rudnicki2012,angulo1}). The corresponding  quantities for the momentum-space probability density $\gamma(\vec{p})$ will be denoted by $R_{q}\left[ \gamma \right]$. The (translationally invariant) Fisher information of the $D$-dimensional density $\rho(\vec{r})$ is defined by
\begin{equation}\label{eq:fisher}
F \left[ \rho \right] := \int_{\Delta} \frac{|\vec{\nabla}_{D}\rho(\vec{r})|^{2}}{\rho(\vec{r})}\, d\vec{r},
\end{equation}
where ${\vec{\nabla}}_D$ denotes the $D$-dimensional gradient of the particle. The corresponding  quantity for the momentum-space probability density $\gamma(\vec{p})$ will be denote by $F \left[ \gamma \right]$. This entropic quantity is playing an increasing role in numerous fields \cite{frieden}, particularly for many-electron systems, partially because of its formal resemblance with kinetic \cite{luo02, frieden} and  Weisz\"acker \cite{romera} energies. The Fisher information, contrary to the Rényi, Shannon and Tsallis entropies, is a local measure of spreading of the density $\rho(\vec{r})$ because it is a gradient functional of $\rho(\vec{r})$, so that it is very sensitive to the density fluctuations. Moreover, these entropic measures allow us to characterize and identify some relevant quantum phenomena such as e.g., quantum phase transitions \cite{nagy2012}, fractality \cite{kinstner}, machine learning \cite{lau2017} and the spectral avoided crossings in atoms and molecules \cite{rosario1,rosario2}. To a great extent this is because the entropic measures satisfy various relevant mathematical properties \cite{hall1999,dembo,sanchez2011} and the position-momentum uncertainty relations \cite{zozor2008,sanchez2011b,rudnicki2012} (see also \cite{sanchez,dehesa2012,portesi}) given by
\begin{equation}
\label{eq:shannon}
S[\rho] + S[\gamma]  \geq  D(1+\log\pi ).
\end{equation}
\begin{equation}
 \label{eq:RUS2}
R_{q}[\rho]+R_{p}[\gamma] \geq D\log\left(p^{\frac{1}{2(p-1)}}q^{\frac{1}{2(q-1)}}\pi\right) \quad with \quad \frac 1p+\frac1q=2.
 \end{equation}
 \begin{equation}
\label{HFI6}
F[\rho]\times F[\gamma] \geq 4D^{2}
\end{equation}
for the Shannon, R\'enyi and Fisher cases, respectively, which improve (i.e., are more stringent) and generalize the Heisenberg position-momentum uncertainty relation $\langle r^{2}\rangle\langle p^{2}\rangle \geq  \frac{D^2}4$.

A more complete way to describe the internal order (or disorder) of the multidimensional hydrogenic systems is given by means of the intrinsic complexity measures, which are composed by two information-theoretic measures which quantify simultaneously two facets of the $D$-dimensional quantum density. The two basic complexity measures of this type are the Cr\'amer-Rao measure \cite{angulo,dehesa_1,antolin_ijqc09} defined by
\begin{equation}
\label{eq:3}
C_{CR}[\rho] := F[\rho] \times V[\rho],
\end{equation}
and the Fisher-Shannon complexity \cite{VignFS,angulo_pla08,romera_1} given by
\begin{equation}
\label{eq:2}
C_{FS}[\rho] := F[\rho] \times \frac{1}{2\pi e}e^{\frac{2}{D}S[\rho]}.
\end{equation}
where $F[\rho], V[\rho]$ and $S[\rho]$ 
denote the Fisher information, the variance 
and the Shannon entropy, respectively, mentioned above.\\
Recently, a generalized complexity measure has been introduced which extend these two previous measures: 
the biparametric LMC-R\'enyi \cite{pipek,lopez,lopezr,romera08}, defined as
\begin{equation}
 \label{eq:6new}
\overline C_{\alpha,\beta}[\rho]: = e^{\frac1D(R_{\alpha}[\rho]-R_{\beta}[\rho])}, \quad 0<\alpha <\beta<\infty, \quad \alpha,\beta \neq 1.
 \end{equation}
Note that the case ($\alpha\rightarrow 1$, $\beta =2$) corresponds to the plain LMC complexity measure \cite{catalan_pre02} $C^{D}_{1,2}[\rho] = \mathcal{D}[\rho] \times e^{S[\rho]}$, which measures the combined balance of the average height of $\rho(\vec{r})$ (by means of the disequilibrium $\mathcal{D}[\rho] = e^{-R_2[\rho]}$) and its total extent over the density support (by means of the Shannon quantity).\\
It is worth to remark that all the previous definitions hold in the momentum space where the radial coordinate, $\vec{r}$, is replaced by the momentum one, $\vec{p}$. These three complexity measures are known to be dimensionless, invariant under translation and scaling transformation \cite{yamano_jmp04,yamano_pa04}, and universally bounded from below \cite{dembo,guerrero,angulo1,lopezrosa} as
\begin{equation}
\label{combounds}
C_{CR}[\rho] \geq D^{2},\quad C_{FS}[\rho] \geq D, \quad 
and \quad \overline C_{\alpha,\beta}[\rho] \geq 1 \quad if \quad \alpha <\beta 
\end{equation}
for $D$-dimensional probability densities. The corresponding complexity measures for the momentum-space probability density $\gamma(\vec{p})$ will be denoted by $C_{CR} \left[ \gamma \right], C_{FS} \left[ \gamma \right]
$ and $\overline C_{\alpha,\beta} \left[ \gamma \right]$, respectively.

  
\section{Hydrogenic and harmonic densities in position and momentum spaces}
\label{sec2} 
 In this section we gather the known probability densities of the $D$-dimensional ($D>1$) hydrogenic and harmonic systems in position and momentum spaces \cite{yanez1994}, which are the basic variables to determine the corresponding uncertainty measures in the rest of the paper. Atomic units are used throughout the paper.
 
\subsection{$D$-dimensional hydrogenic probability density}

The stationary bound states $(n,l,\{\mu\})$ of the $D$-dimensional ($D>1$) hydrogenic system (i.e. a particle subject to a central potential of the form $\mathcal{V}_H(r)= - \frac{Z}{r}$, being $Z$ the nuclear charge) are known (see e.g., \cite{yanez1994}) to have the energies $E_{\eta} = -\frac{Z^{2}}{\eta^{2}}$, (with the grand quantum number $\eta = n+ \frac{D-3}{2}$, and $n=1,2, \ldots$) and the associated probability density is given by
\begin{equation}
\label{eq:21}
\rho^{(H)}_{n,l,\{\mu \}} (\vec{r}) = \frac{\Lambda^{-D}}{2\eta}\frac{\omega_{2L+1}(\tilde{r})}{\tilde{r}^{D-2}}[\tilde{\mathcal{L}}^{(2L+1)}_{\eta-L-1}(\tilde{r})]^{2} |\mathcal{Y}_{l,\{\mu \}}(\Omega_{D-1}) |^{2},
\end{equation}
in position space, where the position vector $\vec{r} = (r, \theta_{1}, \theta_{2},\ldots,\theta_{D-1})$ in polar hyperspherical coordinates, $(n,l,\{\mu \}) = (n,l\equiv \mu_{1}, \ldots, \mu_{D-1})$ are the corresponding hyperquantum numbers with values \{$l= 0,1,2, \ldots, n-1; \quad l\geq \mu_{2} \geq \ldots \geq \mu_{D-1} \equiv |m | \geq 0$\},
and 
\begin{equation}
\label{eq:23}
L \equiv l + \frac{D-3}{2}, \quad \tilde{r} \equiv \frac{r}{\lambda} \quad and \quad \Lambda \equiv \frac{\eta}{2Z}.
\end{equation}
The symbol $\tilde{\mathcal{L}}^{(\alpha)}_{k}(x)$ denotes the orthonormal Laguerre polynomials of degree $k$ and parameter $\alpha = 2l + D-1$ with respect to the weight function $\omega_{\alpha}(x) = x^{\alpha}e^{-x}$ on the interval $[0,\infty)$. The angular part, $\mathcal{Y}_{l,\{\mu \}}(\Omega_{D-1}) $, denotes the hyperspherical harmonics \cite{yanez1994,avery} given by
\begin{equation}
\label{eq:24}
\mathcal{Y}_{l,\{\mu \}}(\Omega_{D-1}) =\frac{1}{\sqrt{2\pi}}e^{im\phi}\prod_{j=1}^{D-2}\tilde{C}^{(\alpha_{j}+\mu_{j+1})}_{\mu_{j}-\mu_{j+1}}(\cos\theta_{j}) (\sin\theta_{j})^{\mu_{j+1}},
\end{equation}
with $\alpha_{j} = \frac{1}{2}(D-j-1)$ and $\tilde{C}^{(\alpha)}_{k}(x) $ denotes the Gegenbauer polynomials of degree $k$ and parameter $\alpha$ orthonormal with respect to the weight function $\omega^{*}_{\alpha}(x) = (1-x^{2})^{\alpha-\frac{1}{2}}$ on the interval $[-1,1]$ \cite{olver} .
\\

Likewise, the probability density of these systems in the $D$-dimensional momentum space \cite{yanez1994} is given as
\begin{equation}
\label{eq:25}
\gamma^{(H)}_{n,l,\{\mu \}}(\vec{p}) = \left(\frac{\eta}{Z} \right)^{D}(1+y)^{3}\left(\frac{1+y}{1-y}\right)^{\frac{D-2}{2}}\omega^{*}_{L+1}(y)[\tilde{C}^{(L+1)}_{\eta-L-1}(y)]^{2}[\mathcal{Y}_{l,\{\mu\}}(\Omega_{D-1})]^{2}
\end{equation}
with $\vec{p} = (p, \theta_{1}, \ldots, \theta_{D-1})$ and the notation
\[
y \equiv \frac{1-\eta^{2}\tilde{p}^{2}}{1+\eta^{2}\tilde{p}^{2}}, \quad and \quad \tilde{p} =\frac{p}{Z}.
\]

\subsection{$D$-dimensional harmonic probability density}

The stationary bound states $(n,l,\{\mu\})$ of the $D$-dimensional harmonic system (i.e. a particle subject to a central potential of the oscillator form $\mathcal{V}_O(r)=\frac{1}{2}\lambda^{2}r^{2}$) are known (see e.g., \cite{yanez1994}) to have the energies $E=\lambda\left(2n+l+\frac D2\right)$ (with $n=0,1,2, \ldots$ and $l=0, 1, 2, \ldots$) and the associated probability density is given by  
\begin{equation}
\label{eq:denspos}
\rho^{(O)}_{n,l,\{\mu\}}(\vec{r}) = \rho^{(O)}_{n,l}(r)\,\, |\mathcal{Y}_{l,\{\mu\}}(\Omega_{D-1})|^{2},
\end{equation}
in position space, where $\rho_{n,l}(r)$ denotes the radial part of the density defined as
\begin{equation}\label{ODRP}
\rho^{(O)}_{n,l}(r)=\frac{2\,n! \lambda^ {l+\frac D2}}{\Gamma\left(n+l+\frac D2\right)} e^{-\lambda r^2}r^{2l}\,\mathcal [\mathcal{L}_{n}^{(l+\frac D2-1)}(\lambda r^2)]^{2},
\end{equation}
with \{$n = 1, 2,3, \ldots; \quad l= 0,1,2, \ldots; \quad l\geq \mu_{2} \geq \ldots \geq \mu_{D-1} \equiv |m | \geq 0$\}. The symbol $\mathcal{L}_{n}^{(\alpha)}(x)$ denotes the orthogonal Laguerre polynomials \cite{olver} with respect to the weight $\omega_\alpha(x)=x^{\alpha} e^{-x}, \, \alpha= l+\frac D2-1,$ on the interval $\left[0,\infty \right)$.\\
On the other hand, in the conjugated space, the Fourier transform provides the following expression
\begin{eqnarray}
\label{eq:momdens}
\gamma^{(O)}_{n,l,\{\mu\}}(\vec{p}) &=& |\tilde{\Psi}^{(O)}_{n,l,\{\mu \}}(\vec{p})|^{2} =  \lambda^{-D}\rho^{(O)}_{n,l,\{\mu\}}\left(\frac{\vec p}{\lambda}\right)
\end{eqnarray}
for the momentum probability density of the $D$-dimensional harmonic stationary state with the hyperquantum numbers $(n,l,\{\mu\})$.\\

\section{Uncertainty relations for hydrogenic and harmonic systems at high $D$}
\label{sec3}
In this section we first realize that the Heisenberg-like uncertainty products of the $D$-dimensional hydrogenic and harmonic systems have the same value at high $D$. Then we prove that the same phenomenon occurs for the other mathematical realizations of the position-momentum uncertainty principle based on uncertainty measures of entropic type, such as the Fisher information and the R\'enyi  and Shannon entropies. 

\subsection{Heisenberg-like uncertainty relation}

For all stationary bound states $(n,l,\{\mu\})$ of both hydrogenic and harmonic systems we have that the position-space variance $V[\rho_{n,l,\{\mu\}}] = \langle \vec{r}^{2}\rangle - \langle \vec{r} \rangle^{2}  =\langle r^{2}\rangle$, since $\langle \vec{r} \rangle= 0$ for any central potential. As well, in momentum space we have that the corresponding variance is $V[\gamma_{n,l,\{\mu\}}] = \langle p^{2}\rangle$. It is known \cite{dehesa2010} that 
$\langle  r^{2} \rangle_H = \frac{\eta^{2}}{2Z^{2}}[5\eta^{2}+1-3L(L+1)]$ and 
	$\langle  p^{2} \rangle_H =	 \frac{Z^{2}}{\eta^{2}}$ for hydrogenic systems, so that the hydrogenic Heisenberg uncertainty product is given as
\begin{equation}
\label{eq:heiH}
\langle r^{2}\rangle_H \langle p^{2}\rangle_H = \frac{D^{2}}{4}\left\{1+\frac{1}{D}(10n-6l-9)+\frac{1}{D^{2}}[10n(n-3)-6l(l-2)+20]  \right\}.
\end{equation}
Similarly, since $\langle  r^{2} \rangle_O = \lambda^{-1}(2n+l+\frac D2)$ and 
	$\langle  p^{2} \rangle_O =	\lambda\left(2n+l+\frac D2\right)$ for harmonic (i.e., oscillator-like) systems, we have the following harmonic Heisenberg uncertainty product 
\begin{equation}
\label{eq:heiO}
\langle r^{2}\rangle_O \langle p^{2}\rangle_O = \left(2n+l+\frac D2\right)^2 = \frac{D^{2}}{4}\left\{1+\frac{1}{D}(8n+4l)+\frac{1}{D^{2}}[4 (2n+l)^2] \right\}.
\end{equation}
Then, it is clear that the Heisenberg uncertainty product has the value 
\begin{equation}
\label{eq:heilike0}
\langle r^{2}\rangle_i \langle p^{2}\rangle_i = \frac{D^{2}}{4} \left(1+ \mathcal{O}\left(\frac{1}{D^{}} \right)\right)
\end{equation}
(with $i = H, O)$ for both $D$-dimensional hydrogenic and harmonic systems. Moreover, it is possible to find that the generalized Heisenberg (or Heisenberg-like) uncertainty product for these two classes of systems is given \cite{toranzo16} by 
\begin{equation}
\label{eq:heilike}
\langle r^\alpha\rangle_i\langle p^\alpha\rangle_i =\left(\frac D2\right)^\alpha \left(1+ \mathcal{O}\left(\frac{1}{D^{}} \right)\right), 
\end{equation}
(with $i=H, O$) which holds for $\alpha \in (-D-2l, D+2l+2)$ in the hydrogenic case and for $\alpha > -D-2l$ in the oscillator case. To obtain this result we have taken into account that the position radial expectation value of the hydrogenic system is 
\begin{eqnarray}
\label{eq:radexpec1}
\langle r^{\alpha} \rangle_H &=& \int r^{\alpha}\rho^{(H)}_{n,l,\{\mu \}}(\vec{r})\, d\vec{r} \nonumber  \\
\label{eq:radexpec1bis}
&=&
 \frac{1}{2\eta}\left(\frac{\eta}{2Z}\right)^{\alpha}\int_{0}^{\infty} \omega_{2l+D-2}(t)[\tilde{\mathcal{L}}^{(2l+D-2)}_{n-l-1}(t)]^{2}\, \tilde{r}^{\alpha+1}\, d\,t \nonumber \\
 \label{eq:radexpec2}
&=&  \left( \frac{D^{2}}{4Z} \right)^{\alpha}\left(1+ \mathcal{O}\left(\frac{1}{D^{}} \right)\right), 
\end{eqnarray}
(which holds for $\alpha > -D-2l$), and the corresponding momentum radial expectation value is 
\begin{eqnarray}
\label{eq:momexpec}
\langle p^{\alpha} \rangle_H &=& \int p^{\alpha}\gamma^{(H)}_{n,l,\{\mu \}}(\vec{p})\, d\vec{p}  \nonumber \\
&=& \left(\frac{Z}{\eta}\right)^{\alpha} \int_{-1}^{1} \omega^{*}_{\nu}(t)[\tilde{\mathcal{C}}_{k}^{(\nu)}(t)]^{2}(1-t)^{\frac{\alpha}{2}}(1+t)^{1-\frac{\alpha}{2}}\,dt\nonumber  \\
\label{eq:momint}
& = & \left(\frac{2Z}{D}\right)^{\alpha}\left(1+ \mathcal{O}\left(\frac{1}{D^{}}\right)\right),
\end{eqnarray} 
which holds for $\alpha \in (-D-2l, D+2l+2)$. Here the notations $k = \eta + L +1 = n-l-1$ and $\nu = L+1 = l + (D-1)/2$ have been used. For the third equality of Eqs. (\ref{eq:radexpec2}) and (\ref{eq:momint}) we have considered that the position and momentum expectation values can be expressed \cite{hey,assche2000,dehesa2010} in terms of hypergeometric functions $_3F_2(1)$ and $_5F_4(1)$, respectively, and then we have used the asymptotics of these functions at high $D$; see \cite{toranzo16} for further details. \\
Similarly, the corresponding radial expectation values of the $D$-dimensional harmonic system are given by 
\begin{eqnarray}
\langle r^{\alpha}\rangle &=& \int r^{\alpha}\rho^{(O)}_{n,l,\{\mu \}}(\vec{r}) d\vec{r} \nonumber\\
&=& \frac{n!\lambda^{-\alpha/2}}{\Gamma(n+l+D/2)}\int_{0}^{\infty} x^{l+\frac{D+\alpha}{2}-1}e^{-x}[\mathcal{L}^{(l+\frac D2-1)}_{n}(x)]^{2}\, dx\nonumber\\
\label{eq:momexpeC}
&=& \left( \frac{D}{2\lambda} \right)^{\frac{\alpha}{2}}\left(1+ \mathcal{O}\left(\frac{1}{D^{}} \right)\right) 
\end{eqnarray}
(valid for $\alpha>-D-2l$) in position space, and 
\begin{eqnarray}
\langle p^{\alpha}\rangle &=& \int p^{\alpha}\gamma^{(O)}_{n,l,\{\mu \}}(\vec{p}) d\vec{p} \nonumber\\
&=& \frac{n!\lambda^{\alpha/2}}{\Gamma(n+l+D/2)}\int_{0}^{\infty} u^{l+\frac{D+\alpha}{2}-1}e^{-u}[\mathcal{L}^{(l+\frac D2-1)}_{n}(u)]^{2}\, du \nonumber\\
\label{eq:mominT}
&=& \left( \frac{\lambda D}{2} \right)^{\frac{\alpha}{2}}\left(1+\mathcal O\left(\frac{1}{D^{}} \right)\right)
\end{eqnarray}
(valid for $\alpha >-D-2l$) in momentum space. In writing the third equality of Eqs. (\ref{eq:momexpeC}) and (\ref{eq:mominT}) we first realize that both quantities are entropic functionals of Laguerre polynomials, and then we use the recently found asymptotics for these functionals at large parameters \cite{temme2017}; see \cite{david1} for further details. Now, note that the multiplication of the hydrogenic expressions (\ref{eq:radexpec2}) and (\ref{eq:momint}), and the harmonic expressions (\ref{eq:momexpeC}) and (\ref{eq:mominT}) gives rise to the wanted result (\ref{eq:heilike}).\\
From Eqs. (\ref{eq:heiH}) and (\ref{eq:heiO}) we remark that the Heisenberg-like products for both $D$-dimensional hydrogenic and harmonic states do not depend on the magnetic hyperquantum numbers of the states. Moreover, from Eqs. (\ref{eq:heilike0}) and (\ref{eq:heilike}) at first order of the pseudoclassical ($D\rightarrow\infty$) limit both Heisenberg-like products are equal to the saturation value of the position-momentum uncertainty relation $(i.e., \langle r^{2}\rangle\langle p^{2}\rangle \geq  \frac{D^2}4$; see also \cite{angulo1993,zozor,angulo2011,guerrero}) which holds for general $D$-dimensional quan	tum systems. Consequently, the first-order Heisenberg-like products at the pseudoclassical border do not capture the qualitatively different character of the Coulomb and quadratic forces which characterize the hydrogenic and harmonic systems, respectively.

\subsection{Fisher-information-based uncertainty relation}

The Fisher information for an arbitrary bound state $(n,l,\{\mu\})$ of a $D$-dimensional single-particle system subject to a central potential $\mathcal{V}(r)$, with the probability densities $\rho \equiv \rho_{n,l,\{\mu \}} (\vec{r})$ and $\gamma \equiv \gamma_{n,l,\{\mu \}} (\vec{p})$ in the two conjugated spaces, can be expressed \cite{romera2005,sanchez2006} as
\begin{equation}\label{eq:fisherPos}
F\left[ \rho \right]= 4 \left\langle p^2 \right\rangle- 2\left| m \right|(2l+D-2) \left\langle r^{-2} \right\rangle
\end{equation}
in position space, and as
\begin{equation}\label{eq:fisherMom}
F\left[ \gamma \right]= 4 \left\langle r^2 \right\rangle- 2\left| m \right|(2l+D-2) \left\langle p^{-2} \right\rangle
\end{equation}
in momentum space, in terms of the pairs of radial expectation values 
$\left( \left\langle p^2 \right\rangle, \left\langle r^{-2} \right\rangle \right)$ and 
$\left( \left\langle r^2 \right\rangle, \left\langle p^{-2} \right\rangle \right)$, respectively.\\
Let us first apply these expressions to the hydrogenic system. Since $\left\langle p^2 \right\rangle_H =\frac{Z^2}{\eta^2} , \left\langle r^{-2} 
\right\rangle_H=\frac{2Z^2}{\eta^3} \frac{1}{2L+1} , \,\,\textrm{and}\,\, \langle p^{-2}\rangle_H=\frac{\eta^2}{Z^2}\frac{8\eta-3(2L+1)}{2L+1}$, one has \cite{dehesa2010} the following values for the position and momentum Fisher informations 
\begin{eqnarray}
\label{HFI2}
F[\rho^{(H)}] &=& \int \frac{|\vec{\nabla}_{D}\rho^{H}(\vec{r})|^{2}}{\rho(\vec{r})}\, d\vec{r}  =\frac{4Z^{2}}{\eta^{3}}[\eta - |m|],  \\
\label{HFI3}
F[\gamma^{(H)}] &=&  \int \frac{|\vec{\nabla}_{D}\gamma^{H}(\vec{p})|^{2}}{\gamma(\vec{p})}\, d\vec{p} =\frac{2\eta^{2}}{Z^{2}}[5\eta^{2}-3L(L+1)-|m|(8\eta-6L-3)+1],\nonumber \\
\end{eqnarray}
in position and momentum spaces, respectively. Then, at high $D$ these expressions simplify as 
\begin{eqnarray}
\label{HFI4}
F[\rho^{(H)}] & =& \frac{16Z^{2}}{D^{2}}\left(1+\mathcal O\left(\frac{1}{D^{}} \right)\right)\\
\label{HFI5}
F[\gamma^{(H)}] & =& \frac{D^{4}}{4Z^{2}}\left(1+\mathcal O\left(\frac{1}{D^{}} \right)\right),
\end{eqnarray}
for the position and momentum Fisher informations at the pseudoclassical limit.\\
Working similarly for the $D$-dimensional harmonic system one finds from Eqs. (\ref{eq:fisherPos}) and (\ref{eq:fisherMom}) the following values 
\begin{eqnarray}
\label{OFI1}
F[\rho^{(O)}] &=& \int \frac{|\vec{\nabla}_{D}\rho^{O}(\vec{r})|^{2}}{\rho(\vec{r})}\, d\vec{r}  = 4\left( \eta +|m| +\frac{3}{2}\right)\lambda, \\
\label{OFI2}
F[\gamma^{(O)}] &=& \int \frac{|\vec{\nabla}_{D}\gamma^{O}(\vec{r})|^{2}}{\rho(\vec{r})}\, d\vec{r}  = 4\left( \eta - |m| +\frac{3}{2}\right)\lambda^{-1},
\end{eqnarray}
for the Fisher information in position and momentum spaces, respectively. Note that both quantities depend on two hyperquantum numbers, $n$ and $m \equiv \mu_{D-1}$ only. Then, for fixed $n$ and $m$ the asymptotics of the Fisher information at high $D$ turns out to be
\begin{eqnarray}
\label{OFI3}
F[\rho^{(O)}] & =& 2\lambda D\left(1+\mathcal O\left(\frac{1}{D^{}} \right)\right), \\
\label{OFI4}
F[\gamma^{(O)}] & =& \frac{2D}{\lambda}\left(1+\mathcal O\left(\frac{1}{D^{}} \right)\right),
\end{eqnarray}
in position and momentum spaces, respectively.\\
Finally, the multiplication of the hydrogenic expressions (\ref{HFI4}) and (\ref{HFI5}) and the harmonic expressions (\ref{OFI3}) and (\ref{OFI4}) gives rise to the following Fisher-information-based uncertainty product 
\begin{equation}
\label{OFI5}
F[\rho^{(i)}]\times F[\gamma^{(i)}] = 4D^{2} \left(1+\mathcal O\left(\frac{1}{D^{}} \right)\right),
\end{equation}
(with $i = H, O$) for both $D$-dimensional hydrogenic and harmonic systems at high $D$. Note that this Fisher-information-based uncertainty product (i) fulfils the Fisher-information-based uncertainty relation (\ref{HFI6}) for general systems, and (ii) cannot disentangle between the Coulomb and oscillator-like forces at the pseudoclassical edge.

\subsection{Rényi-information-based uncertainty relation}

The Rényi entropy for a generic $(n,l,\{\mu\})$-state of a $D$-dimensional hydrogenic system is given, according to Eqs. (\ref{REN}), (\ref{eq:21}) and (\ref{eq:25}), by
\begin{equation}
\label{HRE1}
R_{q}[\rho^{(H)}_{n,l,\{\mu \}}] := \frac{1}{1-q}\log \left( \int_{\Delta} [\rho^{(H)}_{n,l,\{\mu \}}(\vec{r})]^{q}\, d\vec{r} \right)	
\end{equation}
and 
\begin{equation}
\label{HRE2}
R_{q}[\gamma^{(H)}_{n,l,\{\mu \}}] := \frac{1}{1-q}\log \left( \int_{\Delta} [\gamma^{(H)}_{n,l,\{\mu \}}(\vec{p})]^{q}\, d\vec{p} \right)	
\end{equation}
in position and momentum spaces, respectively.
These two quantities can be decomposed into two radial and angular parts which can be expressed in terms of entropic functionals of Laguerre and Gegenbauer polynomials, respectively. Then, we can use the 2017-dated asymptotics of these functionals for large values of the polynomial parameters \cite{temme2017} to find the following values for the position and momentum R\'enyi  entropies of the high $D$-dimensional hydrogenic system \cite{david2}:
\begin{eqnarray}
\label{HRE3}
R_{q}[\rho^{(H)}_{n,l,\{\mu \}}]
&=& \frac{3}{2}D\log \left(\frac{D}{2} \right) + D \log\left(\frac{q^{\frac{1}{q-1}}}{Z}\sqrt{\frac{\pi}{e}} \right) +\frac{q(n-l-1)}{1-q}\log D \nonumber \\
& & +\mathcal O(1), \\
\label{HRE4}
R_{q}[\gamma^{(H)}_{n,l,\{\mu \}}]
& =& -\frac{3}{2}D\log\left(\frac{D}{2}\right)+ D\log \left(\frac{Z\sqrt{e\pi}}{\tilde q^{\frac{1}{q-1}}} \right)+\frac{q(n-l-1)}{1-q}\log D \nonumber \\
 & & +\mathcal O(1), 
\end{eqnarray}
respectively, where $\tilde q=\left(\frac{(2q-1)^{2q-1}}{q^{2q}}\right)^{\frac{1}{2}}$. Note that the Coulomb-strength manifestation appears in the second term and the dependence on the quantum numbers $(n,l)$ do not appear up to the third term in both position and momentum R\'enyi entropies. \\
Similar operations in the $D$-dimensional harmonic (oscillator-like) system have led us to the following values for the position and momentum R\'enyi entropies of this system at high $D$:
\begin{eqnarray}
\label{ORE1}
R_{q}[\rho^{(O)}_{n,l,\{\mu \}}] &=& \frac{1}{1-q}\log \left( \int_{\Delta} [\rho^{O}_{n,l,\{\mu \}}(\vec{r})]^{q}\, d\vec{r} \right)	
=\frac{D}{2}\log\left(\frac{q^{\frac{1}{q-1}}\pi}{\lambda}\right)+\frac{qn}{1-q}\log D\nonumber \\
& & + \mathcal O(1)   \\
\label{ORE2}
R_{q}[\gamma^{(O)}_{n,l,\{\mu \}}] &=& \frac{1}{1-q}\log \left( \int_{\Delta} [\gamma^{O}_{n,l,\{\mu \}}(\vec{p})]^{q}\, d\vec{p} \right)	
= \frac{D}{2}\log(q^{\frac{1}{q-1}}\pi\lambda) + \frac{qn}{1-q}\log D \nonumber \\
& & + \mathcal O(1) ,
\end{eqnarray}
in position and momentum spaces, respectively. Note that the oscillator-strength manifestation appears in the dominant term and the dependence on the quantum numbers $(n,l)$ does not appear up to the second term in both position and momentum R\'enyi entropies. \\
Finally, from the hydrogenic expressions (\ref{HRE3}) and (\ref{HRE4}) and the harmonic expressions (\ref{ORE1}) and (\ref{ORE2}) we can determine the position-momentum uncertainty R\'enyi-entropy-based sum for both high-dimensional hydrogenic and harmonic systems, obtaining the values
\begin{equation}
\label{renyiHO}
R_p[\rho^{(i)}_{n,l,\{\mu\}}]+R_q[\gamma^{(i)}_{n,l\{\mu\}}] \sim D\log(\pi p^{\frac1{2(p-1)}}q^{\frac1{2(q-1)}}), \quad if \quad D >> 1,
\end{equation}
(with $i = H, O$) where $\frac 1p+\frac1q=2$ (i.e., $\frac {p}{p-1}+\frac{q}{q-1}=0$ or $q=\frac {p}{2p-1}$), which saturates the uncertainty R\'enyi-entropy-based relation (\ref{eq:RUS2}) in both hydrogenic and harmonic cases. The symbol $A\sim B \Leftrightarrow \frac{A}{B} \rightarrow 1$.\\
Note that in the limiting case with $q\to1$ and $p\to1$, this expression gives the following dominant term of the position-momentum Shannon-entropy-based uncertainty sum  
\begin{equation}
\label{shannonHO}
S[\rho^{(i)}_{n,l,\{\mu\}}]+S[\gamma^{(i)}_{n,l,\{\mu\}}] \sim D\,\log\,(e\pi), \quad if \quad D >> 1,
\end{equation}
(with $i = H,O$) for both high-dimensional hydrogenic and harmonic systems, which saturates the uncertainty Shannon-entropy-based relation (\ref{eq:shannon}) in both hydrogenic and harmonic cases. In doing this, we have taken into account the definition (\ref{SHA}) of the Shannon entropy for the position and momentum probability densities of these systems which are given by Eqs. (\ref{eq:21}) and (\ref{eq:25}), (\ref{eq:denspos}) and (\ref{eq:momdens}), respectively.\\
Finally, from expressions (\ref{renyiHO}) and (\ref{shannonHO}) we observe that the uncertainty R\'enyi-entropy-based and Shannon-entropy-based sums cannot unravel either between the Coulomb and oscillator-like forces at the pseudoclassical border. In conclusion, neither the Heisenberg-like and the Fisher-information-based uncertainty products nor the entropic uncertainty sums are able to capture any force manifestation at the pseudoclassical border, at least at first order. 

\section{Complexities of hydrogenic and harmonic systems at high $D$}
\label{sec4}

In this section we will determine the leading term of the position and momentum complexity measures of Cr\'amer-Rao, Fisher-Shannon and LMC-R\'enyi types for high-dimensional systems of hydrogenic and harmonic character.

The Crámer-Rao complexity (\ref{eq:3}) of the $D$-dimensional hydrogenic and harmonic states characterized by the hyperquantum numbers $(n,l,\{\mu\})$ are given by 
\begin{eqnarray}
\label{eq:CRpos}
C_{CR}[\rho^{(i)}] &=& F[\rho^{(i)}] \times V[\rho^{(i)}] = D^2 + \mathcal O(D),\\
\label{eq:CRmom}
C_{CR}[\gamma^{(i)}] &=& F[\gamma^{(i)}] \times V[\gamma^{(i)}] = D^2 + \mathcal O(D),
\end{eqnarray}
(for $i = H,O$) in position and momentum spaces, respectively.
To obtain these high-dimensional complexity values we have used (a) the expressions (\ref{HFI4}), (\ref{HFI5}), (\ref{OFI3}) and (\ref{OFI4})  for the position and momentum Fisher information of the hydrogenic and harmonic systems, respectively, and (b) the following values for the involved variances
\begin{eqnarray}
\label{OV1}
V^{(H)}[\rho_{n,l,\{ \mu\}}] & \sim & \left(\frac{D^2}{4Z}\right)^2; \quad V^{(H)}[\gamma_{n,l,\{ \mu\}}]  \sim \left(\frac{2Z}{D}\right)^2,\quad if \quad D >> 1,\\ 
V^{(O)}[\rho_{n,l,\{ \mu\}}] & \sim &  \frac{D}{2\lambda};\quad
V^{(O)}[\gamma_{n,l,\{ \mu\}}]  \sim  \frac{\lambda D}{2} ,\quad if \quad D >> 1,
\end{eqnarray}
which were derived from the general related expressions (\ref{eq:radexpec2}), (\ref{eq:momint}), (\ref{eq:momexpeC}) and (\ref{eq:mominT}) found in the previous section.\\ 
From Eqs. (\ref{eq:CRpos}) and (\ref{eq:CRmom}) we realize that these  Cr\'amer-Rao values attain the universal lower bound $D^2$ in both hydrogenic and harmonic cases at the high-dimensional limit. This means that the Cr\'amer-Rao complexity cannot untangle between Coulomb and oscillator-like systems at the pseudoclassical edge.

The Fisher-Shannon complexities (\ref{eq:2}) of the $D$-dimensional hydrogenic and harmonic states characterized by the hyperquantum numbers $(n,l,\{\mu\})$ are given by 
\begin{eqnarray}
\label{eq:FS2}
C_{FS}[\rho^{(i)}] &=& F[\rho^{(i)}] \times \frac{1}{2\pi e}e^{\frac{2}{D}S[\rho^{(i)}]} = D + \mathcal O(1),\\
\label{eq:FS3}
C_{FS}[\gamma^{(i)}] &=& F[\gamma^{(i)}] \times \frac{1}{2\pi e}e^{\frac{2}{D}S[\gamma^{(i)}]} = D + \mathcal O(1),	
\end{eqnarray}
(for $i = H,O$) in position and momentum space, respectively. These high-dimensional complexity values have been obtained by means of (a) the expressions (\ref{HFI4}), (\ref{HFI5}), (\ref{OFI3}) and (\ref{OFI4}) for the position and momentum Fisher information of the hydrogenic and harmonic systems, respectively, and (b) the following values for the involved Shannon entropies \cite{david2} 
\begin{eqnarray}
\label{HS3}
S[\rho^{(H)}] &=& \frac{3}{2}D\log \left(\frac{D}{2}\right) + D\log \left(\frac{\sqrt{e\pi}}{Z}\right)+\mathcal O(\log D),\\	
\label{HS4}
S[\gamma^{(H)}] &=& -\frac{3}{2}D\log \left(\frac{D}{2}\right) + D\log (Z\sqrt{e\pi} )+\mathcal O(\log D),
\end{eqnarray}
and
\begin{eqnarray}
\label{OSE1}
S^{(O)}[\rho_{n,l,\{\mu \}}] &\sim & \frac{D}{2}\log \left(\frac{e\pi}{\lambda}\right) \\
\label{OSE2}
S^{(O)}[\gamma_{n,l,\{\mu \}}] &\sim& \frac{D}{2}\log (e\pi\lambda),
\end{eqnarray}
for hydrogenic and harmonic systems in position and momentum spaces, respectively.\\
Finally we observe from Eqs. (\ref{eq:FS2}) and (\ref{eq:FS3}) that these Fisher-Shannon values reach the universal lower bound $D$ in both hydrogenic and harmonic cases at the high-dimensional limit. This means that the Fisher-Shannon complexity cannot disentangle either between Coulomb and oscillator-like systems at the pseudoclassical border.\\ 
The LMC-R\'enyi complexities (\ref{eq:6new}) of the $D$-dimensional hydrogenic and harmonic states characterized by the hyperquantum numbers $(n,l,\{\mu\})$ are given by 
\begin{eqnarray}
\overline C_{\alpha,\beta}[\rho^{(i)}] &=& e^{\frac1D(R_{\alpha}[\rho^{(i)}]-R_{\beta}[\rho^{(i)}])},\\
\overline C_{\alpha,\beta}[\gamma^{(i)}] &=& e^{\frac1D(R_{\alpha}[\gamma^{(i)}]-R_{\beta}[\gamma^{(i)}])}, 
\end{eqnarray}
(with $i = H, O$) in position and momentum space, respectively, for $0<\alpha <\beta<\infty$ and $\alpha,\beta \neq 1$.

The determination of the these quantities at high dimensions is straightforward but a bit tedious from the high-dimensional expressions of the corresponding position and momentum R\'enyi entropies which were derived in the previous section. Indeed, from the expressions (\ref{HRE3}) and (\ref{ORE1}) in position space and the expressions (\ref{HRE4}) and (\ref{ORE2}) in momentum space we have obtained the following dominant term for the hydrogenic and harmonic LMC-R\'enyi complexity values 
\begin{eqnarray}
\label{HLMCR3}
 \overline C_{\alpha,\beta}[\rho^{(H)}_{n,l,\{\mu \}}] &\sim  &
\left(\frac{\alpha ^{\frac{1}{\alpha -1}}}{\beta ^{\frac{1}{\beta -1}}}\right)\\
\label{OLMCR1}
 \overline C_{\alpha,\beta}[\rho^{(O)}_{n,l,\{\mu \}}] 
& \sim& \left(\frac{\alpha ^{\frac{1}{\alpha -1}}}{\beta ^{\frac{1}{\beta -1}}}\right)^{\frac{1}{2}}
\end{eqnarray}
in position space, and 
\begin{eqnarray}
\label{HLMCR4}
 \overline C_{\alpha,\beta}[\gamma^{(H)}_{n,l,\{\mu\}}]&\sim &
  \left(\frac{(2\alpha-1)^{\frac{2\alpha-1}{2(1-\alpha)}}}{(2\beta-1)^{\frac{2\beta-1}{2(1-\beta)}}}\right)\left(\frac{\alpha^{\frac{\alpha}{\alpha-1}}}{\beta^{\frac{\beta}{\beta-1}}}\right)\\ 
 \label{OLMCR2}
\overline C_{\alpha,\beta}[\gamma^{(O)}_{n,l,\{\mu \}}] 
& \sim & \left(\frac{\alpha ^{\frac{1}{\alpha -1}}}{\beta ^{\frac{1}{\beta -1}}}\right)^{\frac{1}{2}},  
\end{eqnarray}
in momentum space in the pseudoclassical limit. We realize by looking at the position expressions (\ref{HLMCR3}) and (\ref{OLMCR1}) and the momentum expressions (\ref{HLMCR4}) and (\ref{OLMCR2}) that, contrary to the uncertainty relations and entropic and complexity quantities previously considered in this work, both the position and momentum LMC-R\'enyi complexities are able to disentangle between the hydrogenic and harmonic systems at first order in the high dimensional limit. Moreover, at this dominant order these two position and momentum generalized measures of complexity do not depend on any hyperquantum numbers which characterize the system's state.

\section{Conclusions}

In this work the $D$-dimensional hydrogenic and harmonic (oscillator-like) systems, which are the main prototypes in the multidimensional quantum physics, have been investigated at the pseudoclassical limit of high dimensions in both position and momentum spaces. We have used various spreading quantities of their associated probability densities, such as the Heisenberg-like measures (radial expectation values of arbitrary orders) and the entropic measures of Fisher, Shannon and R\'enyi types. As well we have studied the corresponding uncertainty relations and the complexity measures of Cr\'amer-Rao, Fisher-Shannon and LMC-R\'enyi types.\\ 
We have found a number of relevant results at first order of the pseudoclassical limit. Let us just highlight that (a) none of the hydrogenic and harmonic spreading quantities have any dependence on the state's hyperquantum numbers, (b) the hydrogenic and harmonic Heisenberg and entropic uncertainty relations based on Fisher, Shannon and R\'enyi entropies saturate the corresponding general uncertainty relations, and (c) the hydrogenic and harmonic Cr\'amer-Rao and Fisher-Shannon complexities have the same values in the two conjugated spaces and attain their universal lower bounds. Then, neither the uncertainty relations nor these complexity measures are able to discern between the dominant forces of Coulomb and quadratic character which characterize the hydrogenic and harmonic systems, respectively. Finally, most interesting, the plain LMC and the generalized LMC-R\'enyi complexity measures are the only ones which can disentangle between the hydrogenic and harmonic systems already at first order in the high dimensional limit.

\section*{Acknowledgments}
This work was partially supported by the grants FQM-7276 and FQM-207 of the Junta de Andaluc\'ia and the MINECO-FEDER (Ministerio de Economía y Competitividad, and the European Regional Development Fund) grants FIS2014-54497P and FIS2014-59311P. The work of I. V. Toranzo was financed by the program FPU of the Spanish Ministerio de Educación.

\end{document}